\def\etal{\it et al. \rm}
\def\simlt{\hbox{ \rlap{\raise 0.425ex\hbox{$<$}}\lower 0.65ex\hbox{$\sim$} }}
\def\simgt{\hbox{ \rlap{\raise 0.425ex\hbox{$>$}}\lower 0.65ex\hbox{$\sim$} }}

\def\msun { \rm {M_\odot}}
\def\rsun{ \rm {R_\odot}}

\def\zbar{{\bar z}}

\def\uthresh{u_{\rm th}}
\def\ucrit{u_{\rm crit}}
\def\umin{u_{\rm min}}
\def\Acrit{A_{\rm crit}}
\def\xprime{x^\prime}
\def\rep{R_p}

\documentstyle[aaspp4,epsf,astrobib]{article}
\begin{document}

\title{ The use of high magnification microlensing events in discovering
extra-solar planets}

\author{
	  Kim Griest and Neda Safizadeh
	}
\begin{center}
{\bf Physics Department, University of California, San Diego, CA 92093}\\
\today 
\end{center}

\begin{abstract} 
Hundreds of gravitational microlensing events have now been detected towards
the Galactic bulge, with many more to come.  
The detection of fine structure in these events
has been theorized to be an excellent way to
discover extra-solar planetary systems along the line-of-sight to the
Galactic center.
We show that by
focusing on high magnification events the probability of detecting
planets of
Jupiter mass or greater in the lensing zone (.6 -1.6 $R_E$) is 
nearly 100\%, with the probability remaining high down
to Saturn masses and substantial even at 10 Earth masses.
This high probability allows a nearly definitive statement to made about the
existence of lensing zone planets in each such system that 
undergoes high magnification.
One might expect
lightcurve deviations caused by the source passing near 
the small primary lens caustic to
be small due to the large distance of the perturbing planet, but this effect
is overcome by the high magnification.
High magnification events are relatively rare (e.g. $\sim 1/20$th
of events have peak magnifications greater than 20), 
but they occur regularly and the peak can be predicted in advance, allowing 
extra-solar planet detection with a relatively
small use of resources over a relatively small amount of time.

\end{abstract}


\section{Introduction}
Microlensing has become a useful tool in astronomy for discovering and
characterizing populations of objects too faint to be seen
by conventional methods.  By repeatedly monitoring millions of stars several
groups have now detected the rare brightenings that occur when a dark
object passes between the Earth and a distant source star
\cite{nat93,eros93,ogle93,duo}.
These detections have now become routine with hundreds of events 
reported towards the Galactic bulge, mostly by the MACHO collaboration
\cite{bulge45,alert}.
The reliable detection of large numbers of 
such lensing events allows one to use them for several auxiliary purposes.
For example, relatively rare microlensing ``fine structure" events, where
deviations from the simple brightening formula 
\cite{pac86,griest91}
are apparent, can be searched for.
These have allowed several new effects to be observed, such
as parallax motion
\cite{gouldparallax,gouldparallax2,parallax}, 
the finite size and proper motion of 
the source star
\cite{finite}, and binary lensing
\cite{maopac,oglebinary,pratt}.

Here we consider the special case of binary lensing when one (or more) of
the companions is actually a planet orbiting the primary lens.
This possibility has been investigated by several groups starting with
\citeN{maopac} and \citeN{gouldloeb}.
They found the remarkable result that detectable fine structure occurs 
relatively frequently even for rather low mass planets.
For example, 
\citeN{gouldloeb}
find for a Jupiter mass planet 5 AU from a solar mass star
the probability of detecting the fine structure caused by the jupiter
is about 17\%, while for a Saturn-like planet the probability is about 3\%. 
These relatively high probabilities occur when the planet is in the
``lensing zone" to be discussed later, but they imply that many planetary
systems could be discovered if a systematic search for microlensing fine
structure were made.  The lightcurve deviations caused by a planet last
only a few hours or days (depending upon the mass of the planet) and
can occur at any time during the much longer ($\sim 40 $ days)
primary lensing event.  
In order to not miss these short excursions, 
round-the-clock monitoring would be required, implying dedicated
telescopes at several locations. 
In return, dozens to hundreds of planetary
detections could be made, more than by other proposed detection methods.
Thus microlensing may be the best way to gather statistics on the frequency,
mass distribution, and semi-major axis distribution of planets.
Microlensing is also sensitive to planetary systems throughout the Galaxy
and not just in the solar neighborhood as are most other planet search
techniques.  The main disadvantage to microlensing is that further
study of individual systems is probably impossible.

Following the early work, contributions have been made by several other groups.
\citeN{bolatto} calculated detection
probabilities, \citeN{bennettrhie} and \citeN{wambs}
extended to Earth mass planets by including the finite source effect,
\citeN{gouldgaudi} discussed extraction of physical parameters from 
observational data, \citeN{tytler}, \citeN{peale}, and \citeN{sackett} 
calculated the number of expected detections for
realistic observing strategies.

\section{Microlensing formulas, caustics, and magnification maps}

Microlensing occurs when an intervening stellar mass
lens passes close to the line-of-sight
between an observer and a distant source star.
For Galactic distances, if the lens is a 
single point-mass, two images form with a separation
of milli-arcseconds, too small to resolve.  
However, since the sum of the areas of the images is larger
than the projected area of the source, the magnification, which is given
by the ratio of these areas, can be significant.  
When the source lies directly behind the lens, the image becomes a ring
of radius $R_E$, and the magnification theoretically becomes infinite.
Points in the source plane where the magnification is infinite are called
caustics, and the positions of the 
images of these caustics are called critical curves.
For a single lens, the caustic is single point behind the lens,
and the critical curve is the Einstein ring.

The scale of the microlensing effect is set by the Einstein ring
\begin{eqnarray}
\label{eq-Re}
R_E = \left(4 G m_l L \xprime (1-\xprime) \over c^2\right)^{1/2} =  
612 \rsun \left( {m_l \over \msun} 
{L\over {\rm kpc}} \xprime(1-\xprime) \right)^{1/2},
\end{eqnarray}
where $m_l$ is the mass of the primary lens, $L=D_{os}$ is the distance to the
source star, $\xprime= D_{ol}/D_{os}$ is the fractional distance of the lens,
$\rsun$ is the solar radius, and $\msun$ is the solar mass.
Throughout, we will scale all lengths to $R_E$.  For convenience note
1 AU is $214.94 \rsun$.

When the lens consists of two point-like objects, the caustic positions 
and shapes
depend upon the planet to lens mass ratio $q = m_p /m_l$, and the 
projected planet-lens separation, $x_p$.
The primary lens is assumed to reside at the origin, and the
planet is along the positive x-axis at $x_p$ in units of $R_E$.
For arbitrary distances and mass ratios the caustic structure can 
be complicated, but for small values of $q$, and for $x_p$ not
precisely unity, the picture is simple.
The point-like single lens caustic becomes
a tiny wedge-like caustic, still located near $x=0$, 
while one or two new caustics appear depending on the planet position.
For planets far from the lens ($x_p>1$), there is one new caustic, a small
diamond-shaped planetary caustic located on the same side of the lens as
the planet, while for $x_p<1$, two small heart-shaped caustics appear 
close together on the opposite side of the lens.  
As discussed in the appendix, the position of these caustics is
given approximately by $x_c= (x_p^2-1)/x_p$.
Figure~\ref{figcausticwhole}
shows the caustics for the case of 
$q=0.003$, corresponding to a Jupiter mass planet around a $0.3\msun$ star.
Panel (a) is for $x_p=1.3$ and panel (b) is for $x_p=1/x_p=0.769$.
We will call the caustic near $x=0$ the
``central" or ``primary lens" caustic, 
and the other caustics the ``planetary" caustics.

The relative motion of the source, lens-system, and observer can be 
described as the source moving behind a static lens plane described by
the projected positions of the lens, planet, and caustics.
We work in the lens plane throughout and project physical sizes 
such as the source stellar radius into dimensionless numbers in
the lens plane by multiplying by $\xprime$ and dividing by $R_E$.
If the planet orbits the primary lens in a plane other than the 
lens plane, its position
is also projected into the lens plane.

For single-point lenses high magnification events occur when the source comes
near the caustic at $x=0$.  If $u$ is the projected distance of the source from
the lens (in units of $R_E$ in the lens plane), 
the magnification is $A = (u^2+2)u^{-1}(u^2+4)^{-1/2} \sim u^{-1}$ for
$A$ large.  The peak magnification $A_{max}$ occurs at $\umin$, the
distance of closest approach.
The trajectory intersects a circle of radius $\umin$ at its closest
approach, with $\beta$ being the angle between
this intersection point and the positive x-axis.
With the addition of a planet we continue to define high magnification
events as those caused by the source approaching the central caustic.
For planetary mass binary systems, the lightcurve will be very close to that of
a single lens for most of its duration.

Planetary fine-structure in the high magnification case will arise
due to the difference between a point caustic and the wedge-like
central caustic.  If $A \sim 1/u$ as the source approaches the caustic
then the size of the deviation $\delta = dA/A \sim -Adu$, where $du$
is the shift in the caustic position due to the planet.
It is shown in the appendix that the size of the central caustic along
the x-axis is
\begin{eqnarray}
\label{eq-causticsize}
u_c = {q x_p \over (x_p-1)^2},
\end{eqnarray}
and that this formula is invariant under a ``duality" transformation
$x_p \rightarrow 1/x_p$.  
As described in the appendix
this formula is good for $q<<1$ and $x_p$ not near unity.
Thus for high magnification events, we expect deviations of order 
\begin{eqnarray}
\label{eq-deviation}
\delta \sim u_c A.
\end{eqnarray}
This estimate of lightcurve deviation is in rough agreement with the more 
precise calculations of the next sections, and the caustic size estimate
(eq.~\ref{eq-causticsize})
is excellent, as can be seen in Figure~\ref{figcaustics}.  
Figure~\ref{figcaustics} shows close-ups
of the central caustic for various planetary positions and planetary
mass ratios.  The $x_p\rightarrow 1/x_p$ symmetry
is also apparent in these figures.

Especially useful when discussing the detection of planets or
the probability of certain types of events occurring, are magnification
maps of the source plane, projected onto the lens plane.  These are formed
by imagining a source at each point in the plane and calculating the resulting 
image sizes and resulting magnifications.  
Observationally, one measures the microlensing lightcurve,
the apparent brightness of a star as a function
of time, and
this is completely described by a track through this magnification map.
The duration of the event depends upon the size of
the Einstein ring and the relative projected transverse speed 
$v_\perp$ of the lens system.  
We will divide all times by $t_E = R_E/v_\perp$, so the time to cross the
Einstein ring diameter is $\Delta t=2.0$.

Examples of magnification maps and the resulting lightcurves can be found in
\citeN{wambs}, and \citeN{gouldloeb}, and several other places.
Figure~\ref{figmap} shows some example maps and Figure~\ref{figlc} shows
some example lightcurves.

There are several techniques to calculate such maps in practice.
For point-like lenses, the lens equation in complex notation is given 
by (e.g. Witt 1990)
\begin{eqnarray}
\label{eq-genlens}
z_s = z + \sum_j m_j /({\bar z_j}-{\bar z}),
\end{eqnarray}
where $z_j=x_j + i y_j$ are the positions of the point masses in the lens
plane, $z_s = x_s + iy_s$ is the position of the source,
$z=x_i+iy_i$ are the position of the images, and the bar denotes complex 
conjugation.  
We can rescale this equation by dividing all lengths by $R_E$ and all masses
by $m_l$, and specialize to just one planet to get
\begin{eqnarray}
\label{eq-twolens}
z_s = z - 1/\zbar + q/( x_p - \zbar).  
\end{eqnarray}

One sees that the mapping from an image position at $z$ to a source position
at $z_s$ is one-to-one and extremely simple;  however the reverse mapping
requires solving the above equation for $z$, and results in a 5th degree
polynomial in $z_s$ (see e.g. Witt 1990).  
The partial magnifications are given by the Jacobian of the mapping
from the source to lens plane evaluated at the image positions.
\begin{eqnarray}
\label{eq-magi}
A_i = \left(1 - {\partial z_s \over \partial \zbar}
 {\bar {\partial z_s \over \partial \zbar}}\right)^{-1},
\end{eqnarray}
where the sign of $A$ gives the parity of the image,
and in our case ${\partial z_s/\partial \zbar} = 1/z^2 + q/(z-x_p)^2$.
Caustics and critical curves are found as points where $A_i=\infty$.
The total magnification is just the sum of the absolute values,
$ A = \sum_i | A_i|$. 

The most direct way to create a magnification map is to solve for $A$
at each point in the source plane by solving the 5th degree polynomial.   
When the source is inside a caustic there are 5 images, while if the source
is outside the caustics there are two spurious solutions and only 3 images.
In the later case,
one of these images is behind the lens and is very small and the other two
are important.
A complication of this approach occurs when the mass of the planet is
small.  The magnification map varies on scales smaller than
the planetary Einstein
ring $\rep$, which can be small compared to the size of the projected source
star radius.  Thus one must integrate the magnification over the
limb-darkened source profile to get an accurate total magnification.
The caustic structure gives rise to singularities in $A$, which make this
integration tricky.
(See  \citeN{bennettrhie}, \citeN{gouldgau}, \citeN{gouldgaudi}, 
for examples of this approach).  
We have developed computer programs that successfully
implement this approach, but they are rather slow.

Alternatively, one can note the simplicity of the mapping from image to source
plane, and simply cover the image plane densely with ``photons" and
then map them back to their source positions.  
The resulting density of source photons 
is proportional to the ratio of image to source areas, and therefore
proportional to the magnification.  See \citeN{wambs} for an example of
this method.  This method intrinsically incorporates the finite source effect,
since one must bin the source photons.  The bin size is the effective
source size.  To consider a larger source size, or to include a round source
with limb-darkening, one merely convolves the magnification map with
a kernel made from the desired source profile.
We mainly used this method in creating our maps, though
we checked them in various ways using direct solution.

Lightcurves of single lens microlensing are simple smooth curves 
\cite{pac86,griest91}, 
while if a planet exists there can in addition be several 
sharp peaks.  The durations of these peaks typically scale
with $\rep \propto \sqrt{q}$, 
and last only a day or two, or even only a few hours, a time to be
compared with the typical primary lens event duration of 40 days
\cite{bulge45}.  An nice set of examples of both magnification maps
and lightcurves can be found in \citeN{wambs}.
For both maps and lightcurves we plot residuals
\begin{eqnarray}
\label{eq-delta}
\delta = {\Delta A \over A} = {A_{binary} - A_{single} \over A_{single}}.
\end{eqnarray}
In Figure~\ref{figmap} we show some maps for $q=10^{-4}$.
Figure~\ref{figmap}(a) shows $x_p=1.3$ where there
is only one planetary caustic, and Figure~\ref{figmap}(b) shows
$x_p=0.8$ where there are two planetary caustics on the other side
of the central caustic.
The lightcurves in Figure~\ref{figlc} are for $q=0.003$, $x_p=1.5$,
$\umin=0.05$, and various angles of approach.  
Note the relatively simple structure of these high magnification lightcurves
(with exception of the trajectory along the x-axis which also hits
the planetary caustic).

The method of magnification maps lets us investigate the effects of sources
of different radii.  Once a high resolution map is produced it
can be quickly convolved with any of various source sizes and profiles,
and the lightcurves and probabilities recomputed.
For our convolution kernel we use a limb-darkened profile given by
$I(r) = .4 + .6 \sqrt{1+r^2/R_*^2}$, where $R_*$ is the stellar radius,
$r$ is the distance from the center of the star, and $I$ is normalized to
give a total flux equal to the preconvolution flux.
The maps in Figure~\ref{figmap}(a) and \ref{figmap}(b) were convolved with
a kernel of radius $u_*=0.003$, while Figure~\ref{figmap}(c) is for a
source radius of $u_*=0.03$.
Note these radii are in units of the Einstein ring (eq.~\ref{eq-Re})
and projected into the lens plane.
So for example, a typical main sequence bulge star of radius $R_*=3\rsun$
projects to $u_*=0.003$ if the lens is at 4 kpc, and to $u_*=0.0084$
if the lens is at 7 kpc.
The source is assumed to be at 8 kpc and the primary lens to
have $m_l=0.3\msun$ in these examples.
A giant star of radius $10\rsun$ projects to $u_*=0.01$ for $\xprime=0.5$,
and $u_*=0.028$ for $\xprime=0.875$.
The map of Figure~\ref{figmap}(c) is therefore descriptive of
$30\rsun$ star with a lens system at 4 kpc or a $10\rsun$
source with a lens system at 7 kpc.
An interesting feature of Figure~\ref{figmap}(c) is the circular
ring around the central caustic.  There is a jump
in magnification as the limb of the source star crosses the caustic.
However,
as the star covers more of the region around the caustic, a cancellation
occurs since there is a negative deviation on one side of the caustic
and a positive deviation on the other.

In Figure~\ref{figconlc} we show the result of increasing source size 
on the deviation lightcurves.  
The lightcurves are all for $q=10^{-4}$, $x_p=1.3$, $\umin=0.02$,
and $\beta=50^0$.
As expected
\cite{gouldfinite,nemiroff,wittmao,finite}, as the source radius increases
the amplitude of the signal decreases and the duration of the deviation
increases.  When the star radius is increased to $u_*=0.03$, then it actually
crosses the central caustic giving rise to two bumps on
the lightcurve.  
These bumps occur when the trajectory crosses the ring seen in 
Figure~\ref{figmap}(c).
If detected, these bumps are very useful since the star is
in effect resolved, and the time between the bumps 
allows the projected transverse velocity to be measured.
For high magnification events the width (and height)
of these bumps are determined
by the size of the central caustic and so information about
$q$ and $x_p$ can also be gleaned.  
In Figures~\ref{figmap} and \ref{figconlc}, however, the width of the ring
and the bumps are not determined by the size of the central caustic, 
but by the resolution of our underlying magnification map.
Extraction of planetary parameters from high magnification
events will be discussed in more detail elsewhere \cite{griestsafi}.

\section{Detecting Planets}
Current experiments to search for planets and other microlensing
fine structure piggy-back off the
very successful MACHO collaboration survey alert 
system \cite{alert,pratt}.  
The MACHO 
collaboration monitors millions of stars each night and checks them
for microlensing.  When a candidate microlensing event is detected,
an alert is sent by email to any interested party
[{\tt http://darkstar.astro.washington.edu; macho@astro.washington.edu}].
Two main follow-up collaborations are underway:  The MACHO GMAN
collaboration \cite{alert,pratt}, and the PLANET collaboration \cite{planet}.
GMAN has detected parallax events, the finite source size and proper
motion fine structure, as well as several binary lens events.
PLANET has followed many events and detected much fine structure as well.
Two new survey systems, EROS II and OGLE II, soon plan to generate alerts,
and new additions to the follow-up networks should make coverage of the short
duration planetary deviations more complete.

Several groups have now calculated the 
probabilities that a well-monitored microlensing alert will give rise
to planetary fine structure.  
In calculating probabilities, workers
calculated typical lightcurves caused by
planetary systems and then defined a detection statistic.
For example, \citeN{maopac} and \citeN{bolatto} defined ``detectable"
as at least one lightcurve point inside an area around the planetary caustic.
\citeN{gouldloeb} considered a planet detectable if any point on
the lightcurve deviated by more than 5\% from the single lens case.
\citeN{bennettrhie} defined detectable as the lightcurve deviating by more
than 4\% for a period of $t_E/200$.
Gould \& Loeb found that for a Jupiter mass planet
at a distance of 5 AU from its sun, the probability of detection thus defined
was nearly 17\%.  Given that Jupiter's mass is $0.001 \msun$, it 
was remarkable
that the detection probability was so high.  They explained this in terms
of ``resonant" lensing, which occurs
when the planet is near the Einstein ring radius
($x_p \approx 1$), and thus discovered the ``lensing zone".  
As discussed in detail in the appendix we 
define the lensing zone as the range $0.618 \leq x_p \leq 1.618$.
For a Saturn mass planet their detection
probability dropped to 3\%, and for smaller
mass planets was even smaller.  

Though they performed a very complete calculation, 
Gould \& Loeb made various approximations.  For example, 
they did not calculate the deviation from a magnification map, but
approximated the region of 5\% deviation as a long rectangular box
in the source plane.  They did not include finite source effects, which should
not be large for the Jupiter and Saturn mass planets upon which
they concentrated, but which could be large for Uranus or Earth mass planets.

By including finite source size effects
\citeN{bennettrhie} and \citeN{wambs} continued the calculation to lower mass
planets where the finite size of the source star can be important.
Bennett \& Rhie found probabilities of 
about 2\% for masses as low as Earth mass.
\citeN{peale}, \citeN{tytler}, and \citeN{sackett} calculated
in detail the number of expected planetary detections for several realistic
observing scenarios, and now several groups are undertaking extensive
microlensing searches for planets.  See \citeN{peale}, \citeN{sackett}, or 
\citeN{sahu} for reviews.
The basic plan is to monitor continuously all bulge stars undergoing
microlensing with the hope of finding planetary signals in a few percent of 
them.

\section{High magnification events}

Planetary magnification
maps have their most pronounced deviations from single-lens maps 
near the planetary caustics.  
The size of these caustics scale directly with
the planet-lens mass ratio, and they are located roughly at positions given
by eq.~\ref{eq-xcaustic}.  
Thus the probability of detecting a planet is roughly 
proportional to the angle averaged cross-sectional area of this
region and this is how \citeN{maopac} and \citeN{bolatto} calculated
planetary detection probabilities.
\citeN{gouldloeb} also pointed out that the region of large deviation
continues on a line from the planetary caustic towards the primary lens.

Gould \& Loeb state that in order to get a large deviation the planet
must come near one of the two primary lens images.  This is equivalent
to saying that large deviations occur when the source
is near the planetary caustics.
In this paper, we point out that
for high magnification events, when the source comes very close to the very
small central caustic, 
large deviations from a single lens lightcurve
also occur.  Thus, even though the planet is not near one of the primary
lens images, planet detection can occur.  This is because 
the high magnification
makes the small changes in the central caustic detectable.
In summary, very close to the lens center, the 
difference between the circularly symmetric single-lens caustic
and the tiny wedge-like binary central caustic causes measurable
asymmetries in the lightcurve.
Examples of these caustics are given in Figure~\ref{figcaustics},
and example lightcurves are given in Figure~\ref{figlc}.
Note from Figure~\ref{figlc} that the structure of high magnification
lightcurves is typically simpler than the structure of planetary caustic 
crossing lightcurves.

In order to quantify this effect, we used our magnification maps to 
calculate a large number of lightcurves.  For comparison purposes,
we defined several ``detection criteria".   $P_5$ is the Gould \& Loeb
criteria that at least one point has a deviation of more than 5\% from
the single lens case.  $P_4$ is our analogue of the Bennett \& Rhie
criteria that the event have a time of at least $t_E/200$ 
with more than a 4\% deviation.
As a challenge to observers, we also defined $P_1$, where the planet
is assumed to be detectable if it spends a duration of at least
$t_E/200$ with a deviation from the single lens case of at least 1\%.
Finally, we define $P_\chi$ using a $\chi^2$ statistic.  We define
$\chi_p^2 =\sum \delta_i^2$ where the sum is over all points for
which $u< 0.2$, that is, the total squared deviation for
points during the time when $A \geq 20$.  We define a planet as
detectable if $\chi_p^2 \geq 0.04$, a number set by trial and error to 
correspond approximately to the sensitivity of $P_5$ and $P_4$.
If the photometric measurement errors were $\sigma_i$, 
this value would correspond to a chi-square of $0.04/\sigma_i^2$.

Note that in calculating the deviation,
Gould \& Loeb subtracted a single lens at $x_l=0$
with the primary lens mass unchanged, 
while Bennett \& Rhie subtracted a single lens of
mass $m_l+m_p$ at the center-of-mass position.  
We tried both these subtraction schemes and did not find any significant
difference.
We use the deviation ratio $\delta = (A_{bin}-A_{sing})/A_{sing}$,
since this quantity
has constant magnitude errors as the magnification increases, close to what
happens in a CCD observation.

To investigate high amplification events we took a sample of
events with $\umin\leq \uthresh$, for $\uthresh = 0.1, 0.05, 0.03$, and $0.02$,
corresponding respectively to single-lens magnifications of 
at least 10, 20, 33, and 50.
The quantity $\umin$ is the distance of closest approach of the source to the 
primary lens (in units
of $R_E$).  The maximum magnification $A_{max} \simeq 1/\umin$ 
for $A_{max} \gg 1$.
Given $u_{th}$, 
the probability of an event occurring with $u_{min}\leq u_{th}$
is known a priori to be equal to $\uthresh/\ucrit$,
where it is assumed that every event with primary lens magnification
greater than $\Acrit$ is alerted upon and monitored
($\ucrit = 1$ for $\Acrit=1.34$).  So for example with $\ucrit=1$,
roughly 3\% of monitored events will have $u_{min} \leq 0.03$.

Figures~\ref{figprobjup} through \ref{figproburan01} 
show the results of the probability calculations.
Remarkable is the very high probability for detecting
planets within the lensing zone.  
Figure~\ref{figprobjup} ($q=0.003$) shows a Jupiter mass planet 
around a $0.3\msun$ star. 
Figure~\ref{figprobsat} ($q=0.001$) shows a Saturn mass planet 
around a $0.3\msun$
star, or equivalently a Jupiter mass planet around 1 $ \msun$ star.
Figure~\ref{figproburan003} ($q=10^{-4}$) shows
a 10 Earth-mass planet around a $0.3\msun$ star. 

For $\uthresh=0.02$ or $0.03$, 
and using $P_5$, the least sensitive of our detection criteria,
basically 100\% of Jupiter mass planets would be detected over the entire
lensing zone and substantially beyond it.
As indicated by eq.~\ref{eq-deviation},
the lensing zone probabilities drop as $\uthresh$ increases and therefore
$A_{max}$ decreases, to as low 
as 90\% for $\uthresh=0.05$ and to as low as 80\% for $\uthresh=0.1$.
The $\chi_p^2$ statistic $P_\chi$ performs
similarly or slightly better than $P_4$ and $P_5$ over the entire range.  
If one could use
$P_1$ by detecting 1\% deviations in the lightcurve then the 
detection probabilities would remain near 100\% far beyond the lensing zone
for all values of $\uthresh$.

For Saturn mass planets, Figure~\ref{figprobsat}
shows that probabilities are also near 100\% 
inside the lensing zone, with 
a minimum of 90\% for $\uthresh=0.02$, and
a minimum of 80\% detected for $\uthresh=0.03$.
The drop-off in sensitivity
is quite rapid outside the lensing zone and for larger values of
$\uthresh$, but stays near 100\% at $x_p \approx 1$, and above
40\% even at the edge of the $\uthresh=0.1$ zone.
If one could detect 1\% deviations, then the probability is again 
nearly 100\% over a wide range of $x_p$.

We note a duality invariance $x_p \rightarrow 1/x_p$ in the probability
plots for high magnification events.  The probability of detecting
a planet at $x_p=0.5$ is the same as detecting a planet at $x_p=2$.
This is because the central caustic is almost identical
under this transformation (see eq.~\ref{eq-causticsize} and 
Figure~\ref{figcaustics}).  
The duality symmetry also shows up in the position of the planetary caustics:
$x_p$ and $1/x_p$ give caustics at the same $x_c$ according to 
eq.~\ref{eq-xcaustic} (see Figure~\ref{figcausticwhole}).
This symmetry implies a degeneracy in determining the planet position
from the lightcurve for high magnification events.  
High magnification lightcurves with a planet at $x_p$ will be almost
identical to those with a planet at $1/x_p$ in most cases.
There are also potential degeneracies between planetary mass and distance,
and these will be considered elsewhere \cite{griestsafi}.
See \citeN{gouldgaudi} for an extensive discussion of degeneracies
for planetary caustic events.

Figures~\ref{figproburan003} and \ref{figproburan01} show the probabilities
for 10 Earth-mass planets ($q=10^{-4}\msun$).
From eq.~\ref{eq-causticsize},
we expect the size of the deviations to drop by a factor of 10 from the
$q=0.001$ case, 
so we expect small probabilities when using 
$P_4$ or $P_5$.  Also we expect statistics such as $P_4$ which require
a minimum time above a threshold to lose sensitivity in comparison
with $P_5$ which requires only one deviant point.  These expectations
are born out in 
Figure~\ref{figproburan003} which shows a maximum probability of
of $\sim 80$\%
near $x_p \approx 1$, dropping rapidly even inside the lensing
zone, and probabilities below 1\% at the edge of the lensing zone.  
$P_1$ fares much better, giving probabilities 80\% -- 100\% near
the zone center, dropping to 20\% -- 50\% near the zone edge.

In order to test the effect of the finite source size on our probabilities
we convolved each map with a kernel representing a limb-darkened
source star of various radii, and then recalculated the probabilities.
For $q=0.003$ we found no significant differences with radii up to
$u_*=0.01$ corresponding to a typical giant star 
at a distance halfway to the Galactic Center.
For $q=10^{-4}$, however, the effect is quite apparent, as is shown in
Figures~\ref{figproburan003} and \ref{figproburan01}.
For $u_*=0.01$, the peak $P_5$ or $P_4$ probabilities are less than 35\% with 
a rapid drop even inside the lensing zone.  
In this case, the convolution has caused the maximum deviations due to
the central caustic to drop below 4\%, so that the detections
are not actually ``high magnification" events but rather are 
caused by trajectories
which pass through the planetary caustic region.
This explains the counter-intuitive result that the 
$\uthresh=0.1$ case has a higher probability than the $\uthresh=0.02$ case.
Low values of
$\uthresh$ force the trajectories to pass near the origin, while
higher values include trajectories that are more likely to hit
the planetary caustic. 
If 1\% deviations could be detected the $P_1$ statistic gives
high probabilities even for 10 Earth-mass planets and giant source stars.
The higher probabilities of
Figure~\ref{figproburan003} show that the central caustic is still important
for $u_*=0.003$.  

We note that all the probabilities calculated here are for the projected
lens-planet separation.
To find the probability of detecting a planetary system with a 
given semi-major
axis our probabilities must be averaged over
the possible inclination angles of the planetary system. 
To find the probability of finding a planet of a given mass one must then
average over a distribution of semi-major axes, and also over the density of 
planetary systems along the line-of-sight, taking into account the
variation of $\xprime$.  
See \citeN{gouldloeb} for an example.
This calculation will be presented elsewhere
\cite{griestsafi}, but see Section~6 for a caveat.

\section{Discussion}

High magnification events have both advantages and disadvantages when
compared with ordinary planetary fine structure events.  
One obvious advantage is that
since the source star is highly magnified, more flux is available and
more accurate photometry can be performed.  For example,
events satisfying the $\uthresh=0.03$ or $u_{th}=0.02$
criteria are 3 to 4 magnitudes
brighter during peak magnification, and thus Poisson errors in the photometry
are reduced.
The obvious disadvantage is that high magnification events occur rarely, 
only 2\%-3\% of the time for the above examples,
so the number of such events will be small.  
In some situations, this disadvantage may be somewhat offset since
fewer telescope resources will be needed to perform the follow-up.
Typical groups searching for planets anticipate monitoring dozens of
events per day in a round-the-clock manner since it is never known when
a few-hour-long planetary excursion will take place.  
This requires a world-wide system of dedicated telescopes.
Since the time of a high magnification peak can be predicted well in advance,
a focus on high magnification events would allow concentration of resources
on the most valuable events.  One could make important discoveries
while monitoring only a fraction of the stars over a fraction of the time.
Larger telescopes which allowed rapid rescheduling could more easily
be brought into play if the time needed was small and the potential
payoff large.
Special purpose equipment to reach more sensitive detection thresholds
might be worthwhile deploying if the chances of success were known
to be large.

So, while the continuous monitoring method will obviously give more total
detections, the cost/benefit ratio is better for high magnification
events.  In addition, the high probability of detection results in a high
efficiency experiment and allows nearly definitive statements to be
made on a system by system basis.  For example, using Figure~\ref{figprobjup},
each non-detection in an $A_{max}>33$ event immediately implies there 
is no planet with
a mass equal to or greater than Jupiter in the lensing zone. 
The high efficiency
also allows statistical results to be obtained with fewer actual measurements.
Another potential advantage of high magnification events is the larger
likelihood of a measurable finite source size effect.  In these cases
the projected transverse proper motion can be found and information
about the lens distance ($\xprime$) can be inferred.  This can help break
the degeneracies, described in \citeN{gouldgaudi}, that make determination
of $q$ and $x_p$ difficult.

\section{Lightcurve fitting vs. a priori subtraction}
All probability predictions made to date have used the deviation between
the binary-lens lightcurve and the single lens lightcurve as the signal
to be detected.  
In this paper we followed suit so as to allow comparison of our probabilities
with previous calculations.
In practice, however,  only the observed binary lightcurve (plus
noise) is known.  In order to extract the signal one can subtract
a single lens lightcurve, but one does not know a priori which single
lens lightcurve to subtract.  It must be found by fitting,
and the fit single-lens lightcurve will try to minimize the binary features
and will reduce the signal.  In order to test the size of this effect
we performed a non-linear fit to a single-lens
lightcurve, and then subtracted that lightcurve.
Examples of the residuals from such subtractions are shown in 
Figure~\ref{figfitlc}.
The effect described above is clear.  The chi-squared fitting procedure
produces a single-lens lightcurve which minimizes the largest deviations;
in chi-square fitting it is better to miss many points 
by a little than a few points by a lot.  
When using threshold detection criteria as was done here and as has been 
done by previous workers, the detection probability can be altered.
In the example of Figure~\ref{figfitlc} the peak deviation is above the $P_5$
detection threshold of $0.05$ when using a priori subtraction,
but below even the $P_4$ threshold of $0.04$ when using the fit subtraction.
Thus we counted this event as detectable in our calculations, while
it would not be detectable by these criteria if the fit subtraction was used.
This effect holds not only for high magnification events but for
all planet detection probabilities near detection threshold.
There may be other detection statistics that are more robust to
fitting, and these will be explored elsewhere \cite{griestsafi}.

\begin{center}
\bf Appendix: Planetary caustic positions, the lensing zone,
and central caustic size, 
\end{center}

Consider a single point-like lens at $x=0$ and a source at $x_s$.
The two images occur along the lens-source line at
\begin{eqnarray}
\label{eq-ximage}
x_i = [x_s \pm (x_s^2+4)^{1/2}]/2. 
\end{eqnarray}
where all distances are measured in units of $R_E$.
A negative value of $x_i$ means the
image is on the other side of the lens from the source.

\begin{figure}
\plotone{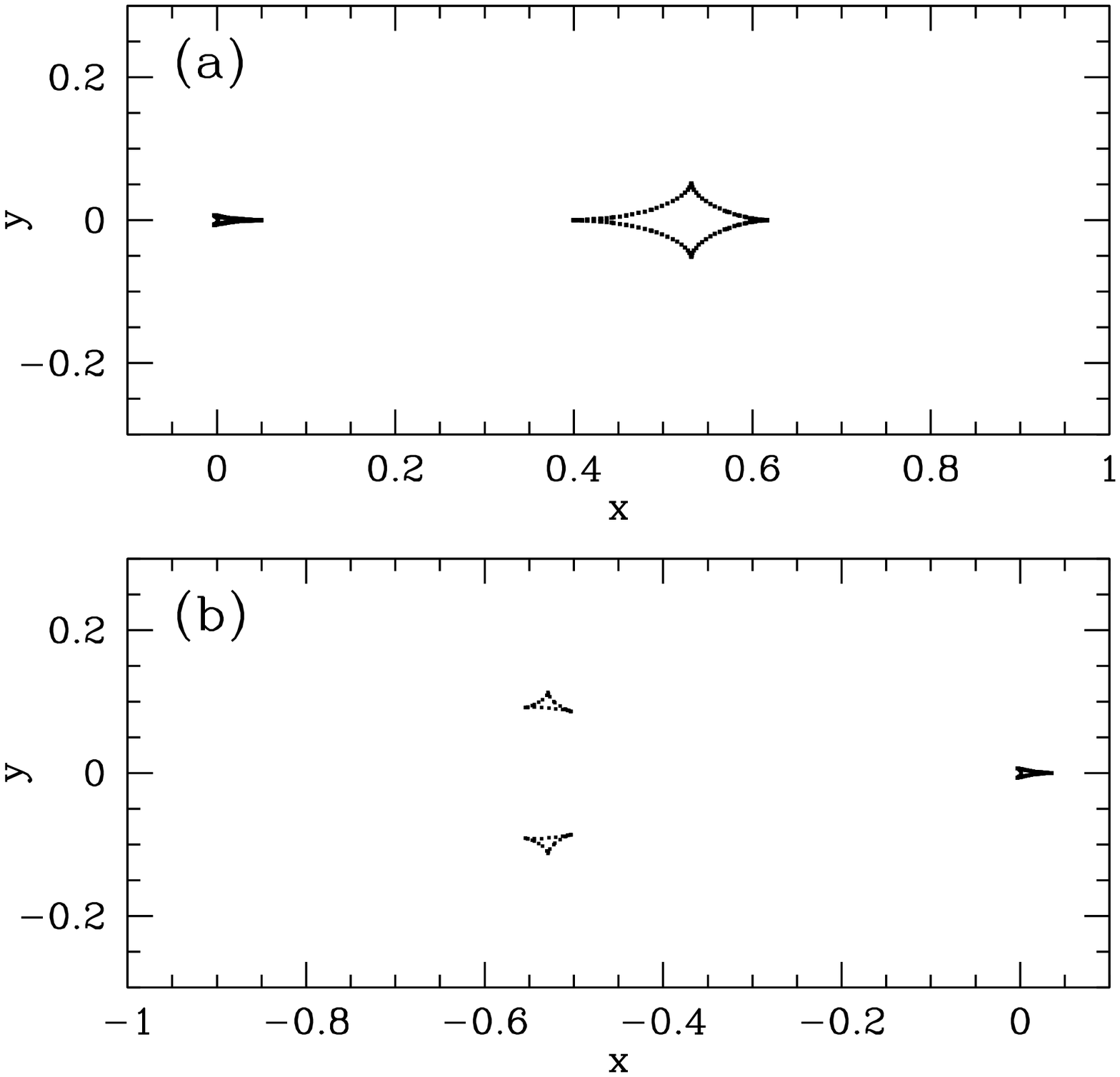}
\caption{
  Caustics for $q=.003$, showing the central primary caustic near
  the origin and the larger planetary caustics.  Part (a) is for
  a planet at $x_p= 1.3$ and part (b) is for the ``dual" position
  at $x_p=1/1.3 = 0.769$.
  \label{figcausticwhole}}
\end{figure}

Since a planet mass is much smaller than the primary lens mass, its
area of influence is small when measured in units of 
$R_E$.  Thus to first approximation the planet can have a large effect
only when its position is near one of the main images ($x_i$).
This is the lens plane point of view.  
From the source plane point of view, one
expects the planet to have a strong effect when the source comes near
the planetary caustics (for example, see Figure~\ref{figcausticwhole}).
Thus the strong effect of the source being near the planetary
caustic is the same as the planet being near one of the single-lens images.
The relation between planet and caustic positions should then be the same
as the relation between image and source positions, that is,
the inverse of eq.~\ref{eq-ximage}.
Thus the caustic position is along the x-axis at
\begin{eqnarray}
\label{eq-xcaustic}
x_{c} \simeq (x_p^2 -1)/x_p, 
\end{eqnarray}
where $x_p$ is the position
of the planet in units of $R_E$.
This formula should work when $m_p \ll m_l$ and $x_p$ not near
unity.  When $m_l \approx m_p$, the planetary influence is no longer small,
and when $x_p \approx 1$, the caustics merge and take complicated shapes.

The ``lensing zone" was first discussed by \citeN{gouldloeb}
as the set of planet-lens distances where the probability of 
detecting the planet was high (see their Figure 4), and has
been used with various definitions by others to mean
the region where the planet is near the Einstein ring.  In searching
for planets one uses as a selection criteria that the primary lens be
magnified by more than some amount such as $A_{th}=1.34$.  
This is because observationally, microlensing is not easy to identify
when the peak magnification is low.
This selection criteria is equivalent to
requiring that the source star pass within some (projected) distance of the
primary lens (e.g. $u_{th} = 1$, for $A_{th}=1.34$).
The probability of detecting the planet is proportional to the averaged
cross-section of some magnification contour in the source plane, 
which is roughly proportional to
the chance that a trajectory that comes within $u_{th}$ 
also hits the planetary caustic.
When the caustic is near the Einstein ring ($x_c \approx 1$)
the probability is high, and
when the caustic is within the ring ($x_c<1$) the probability is also high, but
when the caustic is far outside the ring ($x_c \gg 1$), the probability drops
at least inversely with distance $x_c$.
Thus the lensing zone can be
defined as those positions $x_p$ for which $x_c \leq 1$. 
Using eq.~\ref{eq-xcaustic}, this definition corresponds to 
a lensing zone 
of $0.618 < |x_p| < 1.618$, values also used by \citeN{wambs}.
While the detection probability drops quickly outside the lensing zone, clearly
there will be some probability when the caustic is 
just outside the Einstein ring.  
Also the probability near the edge of the zone will
be smaller than in the middle.  Finally, since the mass of 
the planet determines the caustic size and region of influence, the edges 
of the zone will be a strong
function of the mass of the planet, and also of the detection criteria used.

\begin{figure}
\epsscale{.8}
\plotone{ 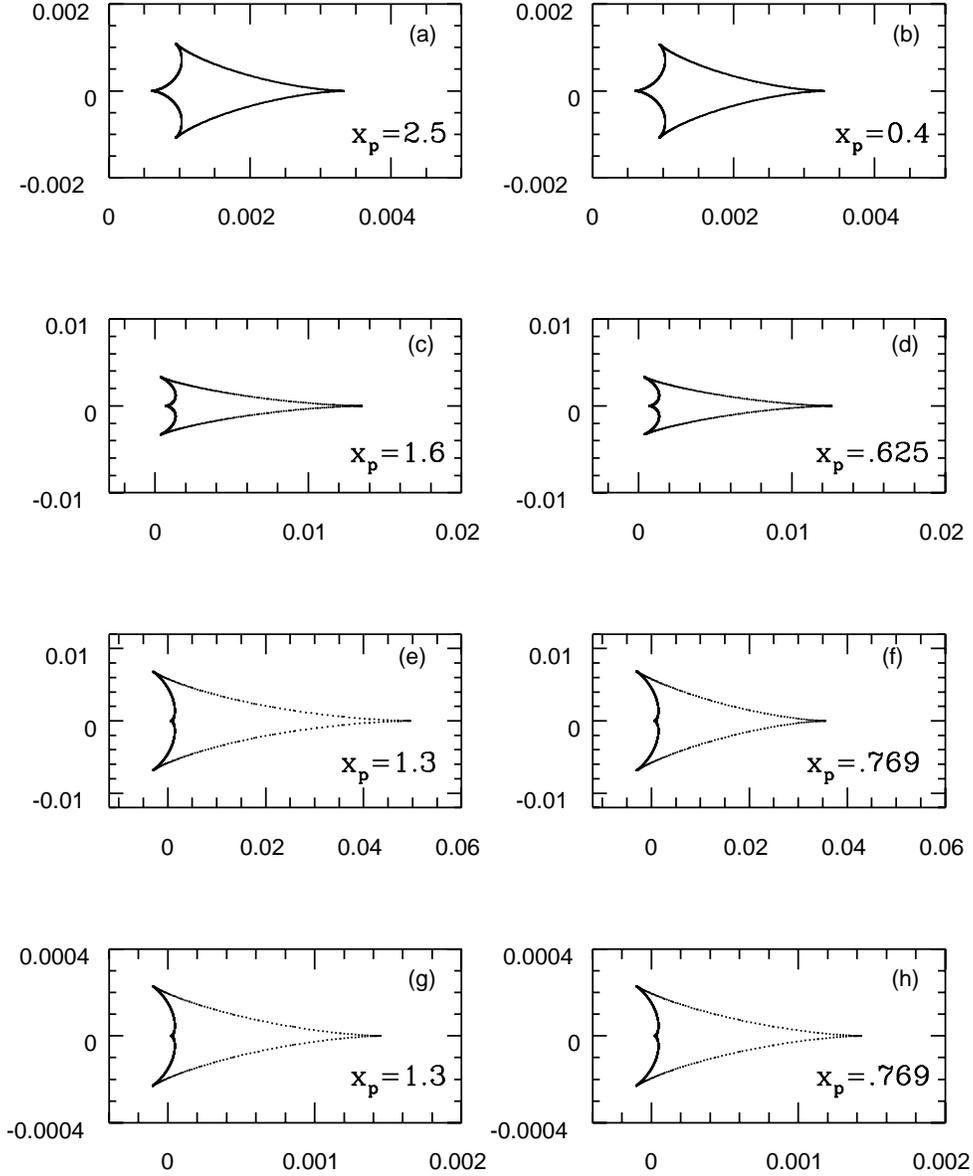}
\caption{
  Close-ups of the central primary lens caustic.  Left-hand panels show
  planetary positions $x_p>1$, while right-hand panels show a planet
  in the dual positions $1/x_p$.  
  All panels except (g) and (h) have $q=.003$.
  Note the excellent match of left and right hand panels except for 
  (e) and (f) where the approximation is just starting to break down.
  Panels (g) and (h) are for $q=.0001$, where the caustics
  are predicted to be 30 times smaller and symmetry restored.
\label{figcaustics}}
\end{figure}

The extent of the central caustic along the x-axis
(Figure~\ref{figcaustics}) can be estimated
as follows.  For a single point-like lens the image of
the point-like caustic is the circular Einstein ring critical curve of
radius 1.  When $q = m_p/m_l <<1$ and $x_p \neq1$, we expect the
binary system critical curve to remain nearly the same and to map
onto the small central caustic (see Figure~\ref{figcaustics}).
The planetary caustics will
map onto one or two small critical curves near the planet.

Restricting ourselves to the x-axis, the planet will affect the central
caustic in two ways.  First, the critical curve that crosses the x-axis
at $x= \pm 1$ in the single lens case, will be moved slightly to 
$x= \pm 1 + \epsilon$.  Second, 
the planet will cause the critical curve image to 
map to a slightly different position on the x-axis.
Eq.~\ref{eq-twolens} says the tip of the central caustic on the x-axis will
occur at $x_s = x -1/x + q/(x_p-x)$.  To find $x$ we find the critical
curve using eq.~\ref{eq-magi} and $A_i=\infty$, or 
\begin{eqnarray}
\label{eq-critical}
1 - \left( {1\over x^2} + {q\over (x-x_p)^2} \right)^2 = 0.
\end{eqnarray}
Let $x=1 + \epsilon$, and solve this equation in the limit of $q \ll 1$
($m_p \ll m_l$) and $\epsilon \ll1$ (critical curve doesn't move much).
This gives
\begin{eqnarray}
\label{eq-epsilon}
\epsilon \simeq {q\over 2 (1-x_p)^2} \left( 1 + {q\over(1-x_p)^3}
\right)^{-1} \simeq {q\over 2 (1-x_p)^2}.
\end{eqnarray}
Inserting $x = 1 + \epsilon$ into eq.~\ref{eq-twolens} gives
\begin{eqnarray}
\label{eq-causticshift}
u_c \equiv  x_s \simeq 2 \epsilon + {q\over x_p-1} 
\simeq {q x_p \over (x_p-1)^2}.
\end{eqnarray}
This is the expected shift from the origin for the caustic tip.
This formula gives a very good prediction of the sizes of the caustics
shown in Figure~\ref{figcaustics}.  
Note the formula is invariant under the transformation
$x_p \rightarrow 1/x_p$ as evidenced in Figure~\ref{figcaustics} 
and displayed in
eq.~\ref{eq-causticsize}.
We expect the formula to break down when $q \rightarrow 1$ or
$x_p \rightarrow 1$.

\acknowledgements

K.G. thanks Jerry Guern for many helpful conversations and work on
early versions of these calculations.
We thank Neal Dalal for 
help on the size of the central caustics, the use
of his caustic finding program, and other useful conversations.  
We thank Art Wolfe for the use of his computer.
We acknowledge
support from an IGPP mini-grant, from the Department of Energy,
from the Alfred P. Sloan Foundation, and from a Cottrell Scholar award
of Research Corporation.


\clearpage

\begin{figure}
\epsscale{.6}
\plotone{ 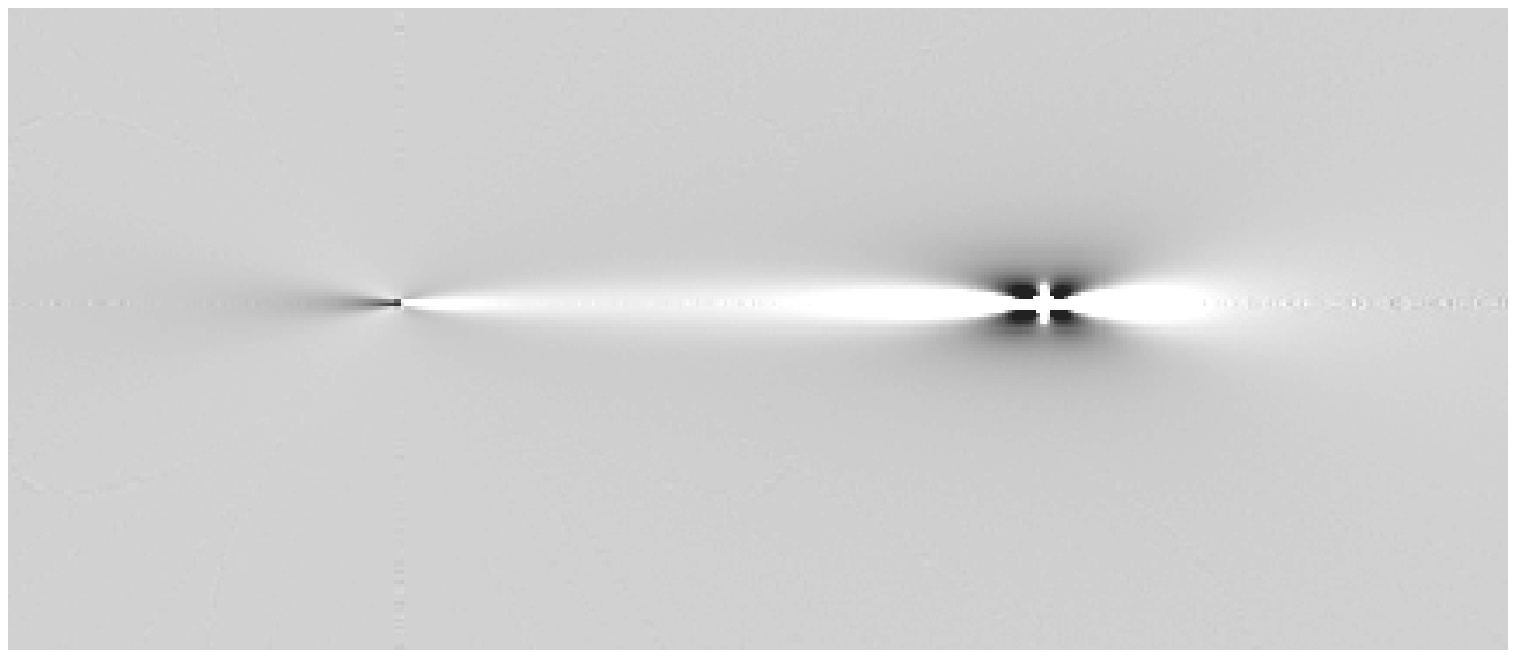}
\end{figure}
\begin{figure}
\epsscale{.6}
\plotone{ 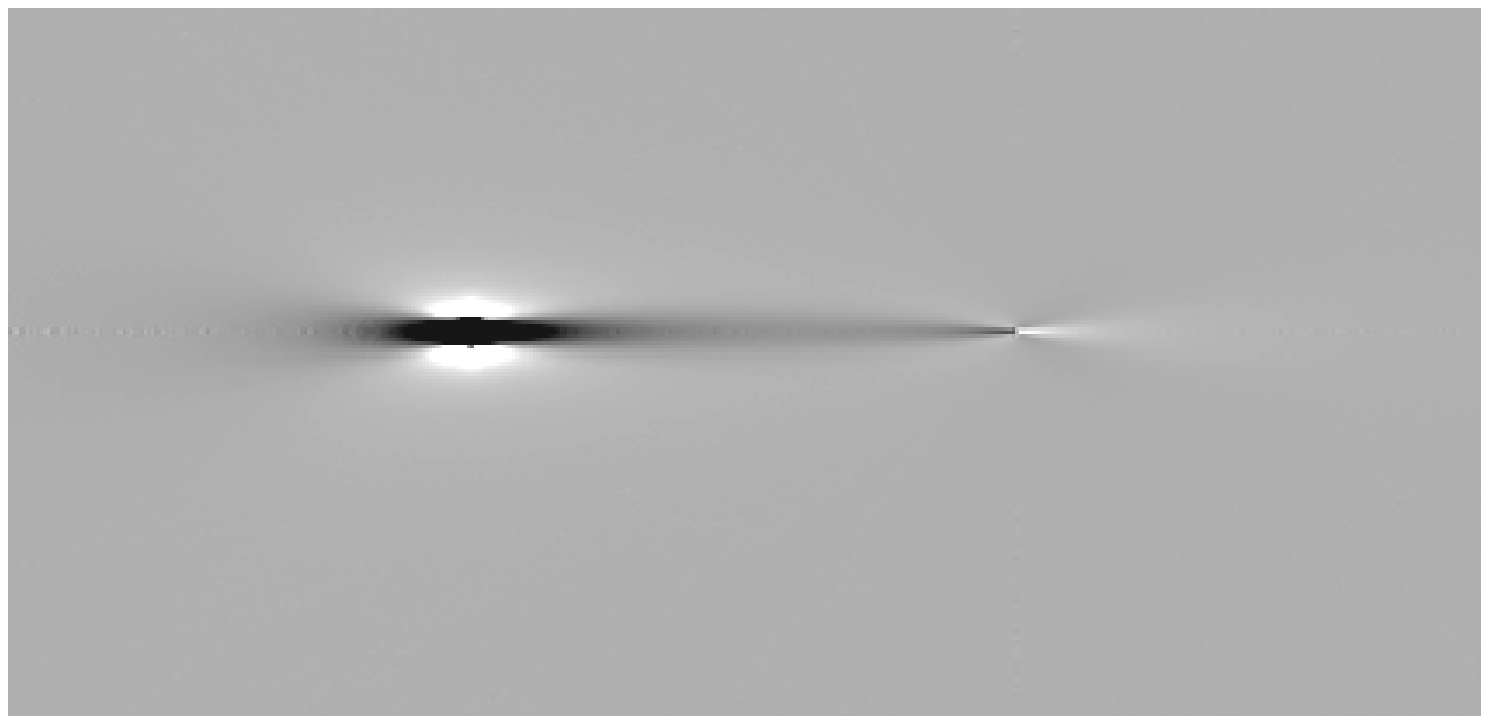}
\end{figure}
\begin{figure}
\epsscale{.6}
\plotone{ 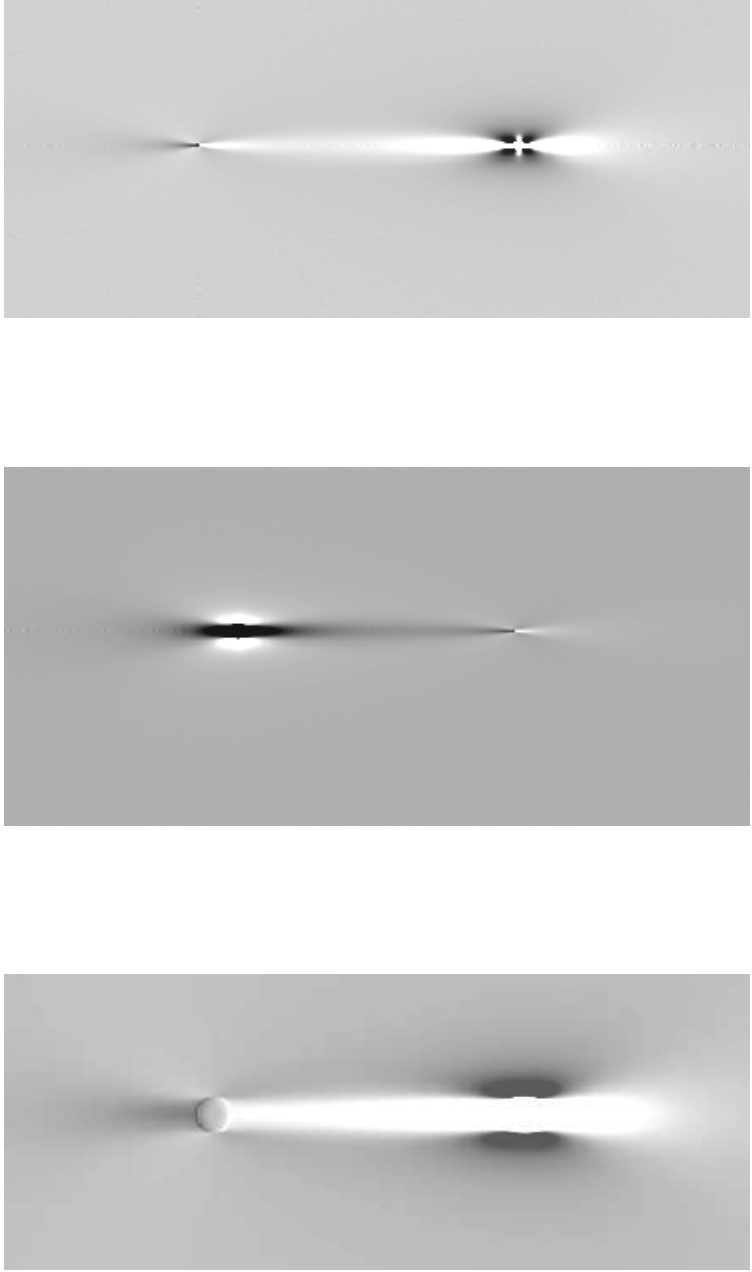}
\caption{
Example magnification maps show the ratio of planetary to
single lens magnification in the source plane.  Light areas show positive
deviations (ratios greater than one) and dark areas show negative deviations.
All panels are for $q=10^{-4}$, a 10 Earth-mass planet around a $0.3\msun$
star. Panel (a) shows $x_p=1.3$ with a source radius of $u_*=0.003$,
while panel (b) shows an example of $x_p<1$ ($x_p=0.8$) 
with the same source radius.
Panel (c) shows $x_p=1.3$, but with the large source radius of $u_*=0.03$.
\label{figmap}}
\end{figure}

\begin{figure}
\epsscale{.8}
\plotone{ 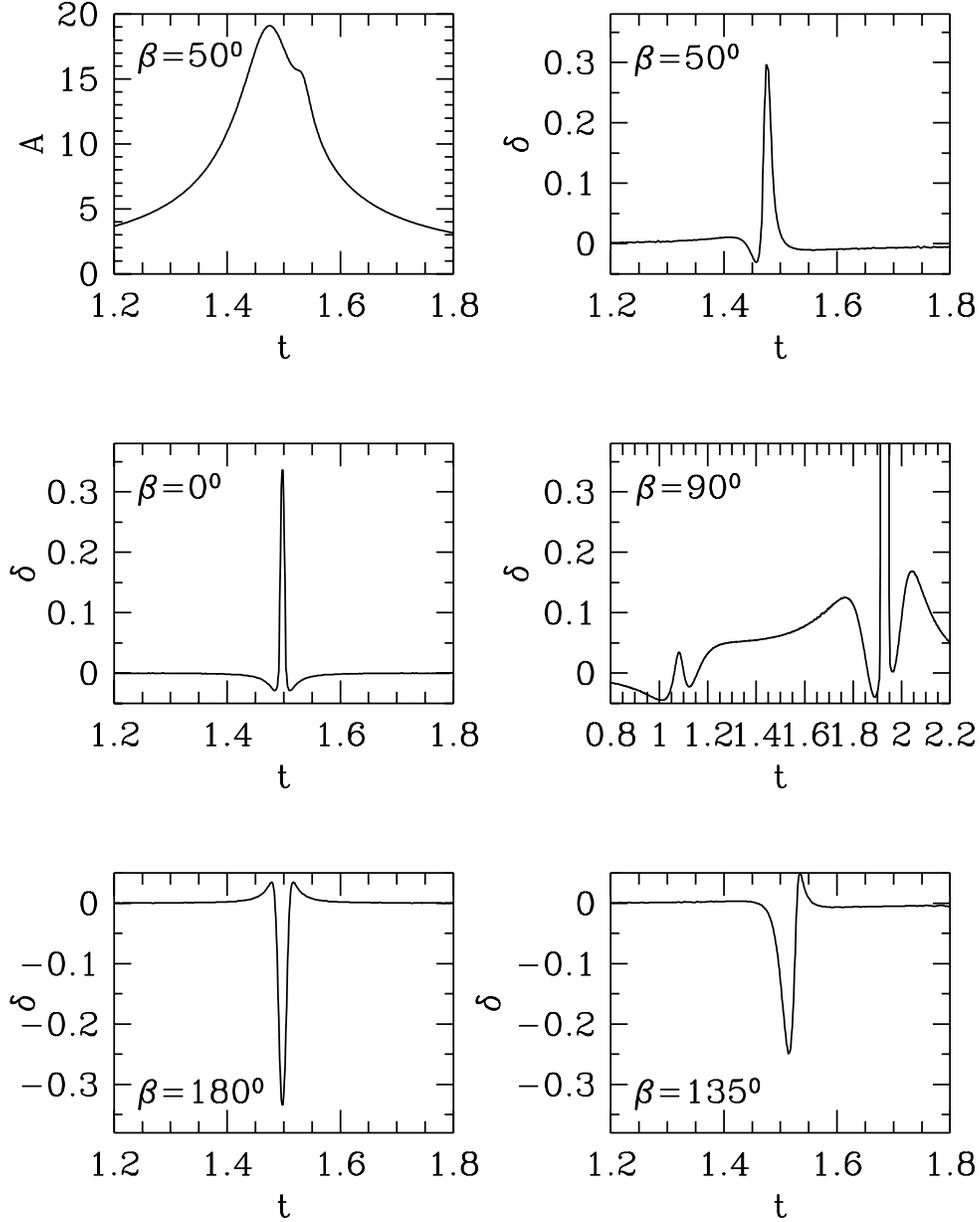}
\caption{
Example high magnification lightcurves, for $q=0.003$, $x_p=1.5$, 
$\umin=0.05$, and various angles of approach to the central caustic.  
The upper left panel shows the total magnification lightcurve, while
the others show only the deviation $\delta = (A_{bin}-A_{single})/A_{single}$.
Time is plotted in units of $t_E$.
The $\beta=0^0$ trajectory is perpendicular to the lens-planet axis 
and between them, while $\beta=180^0$ is on the side opposite the planet.
As is apparent, the $\beta=90^0$ trajectory also hits the 
larger planetary caustic.
\label{figlc}}
\end{figure}

\begin{figure}
\plotfiddle{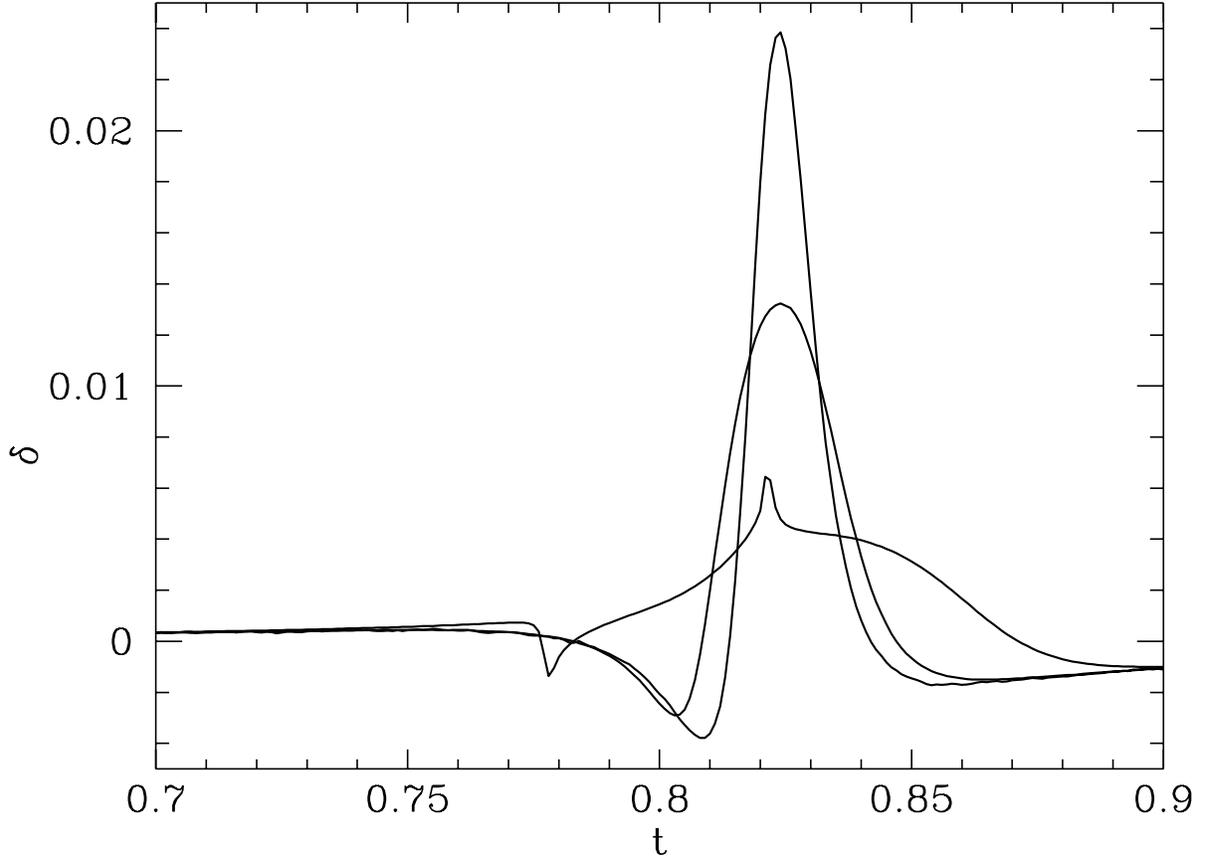}{7cm}{-90}{60}{60}{-230}{360}
\caption{
Comparison of planetary deviation lightcurve for different source radii:
$u_*=0.003$, 0.01, and 0.03, corresponding
to $R_*=3\rsun$ with lens at 4 kpc, $R_*=10\rsun$ with lens at 4 kpc, 
and $R_*10 \rsun$ with lens at 7 kpc, respectively.
The deviation $\delta$ is plotted
vs. the time in units of $t_E$. These curves are for $q=10^{-4}$,
$x_p=1.3$, $\umin=0.02$, and $\beta=50$ degrees.
The smaller the stellar radius, the higher and sharper the deviation.
The bumps
in the $u_*=0.03$ radius curve occur as the limb of the star crosses
the central caustic. 
\label{figconlc}}
\end{figure}

\begin{figure}
\plotfiddle{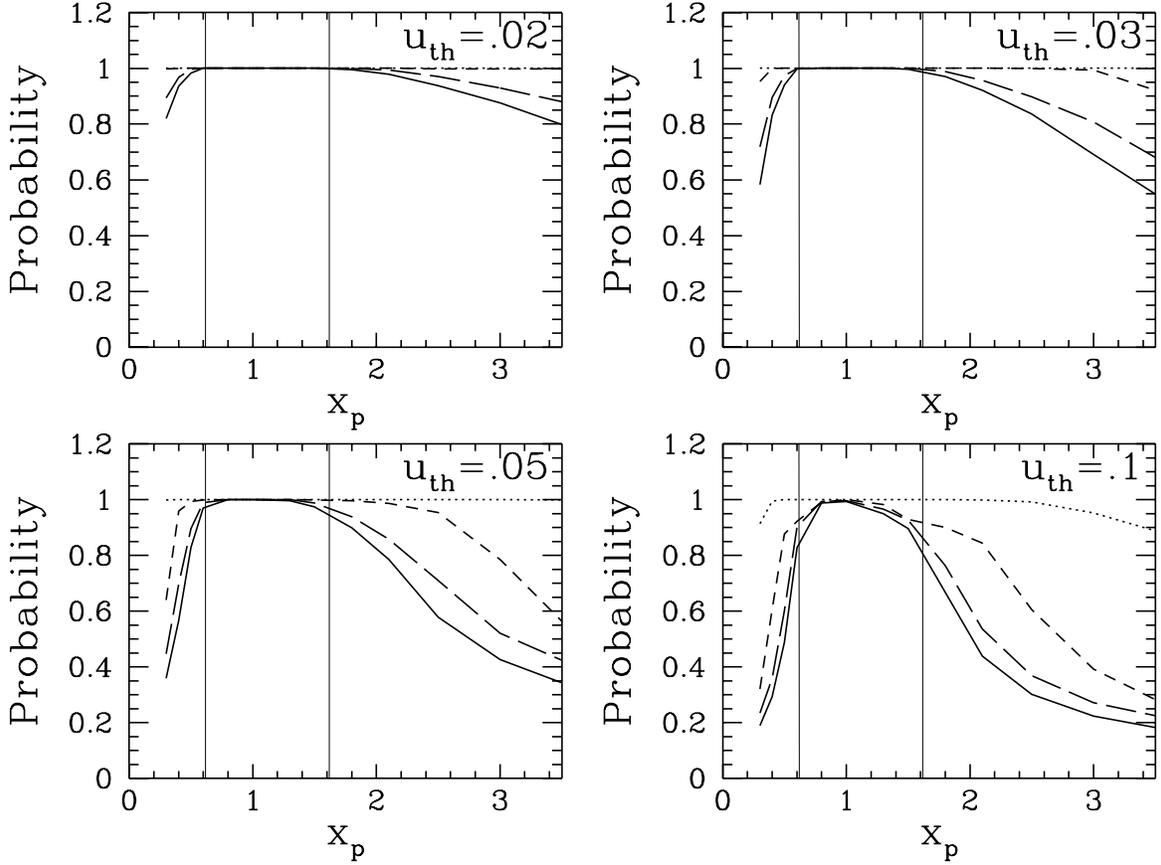}{7cm}{-90}{60}{60}{-230}{360}
\caption{
Probability of planetary detection for high magnification events for
a planet/lens mass ratio of $q=0.003$, corresponding to a Jupiter
mass planet around a $0.3\msun$ star.  The probability is plotted
vs. the planet-lens separation $x_p$ in units of $R_E$.
Each panel shows a different value of threshold $\uthresh$, where
only events which have $\umin \leq \uthresh$ are counted.
Four different detection statistics are plotted in each panel:
$P_5$ is solid line, $P_4$ is long-dash line, $P_\chi$ is short-dash line,
and $P_1$ is dotted line.  The light vertical lines demark the lensing zone.
\label{figprobjup}}
\end{figure}
\clearpage

\begin{figure}
\plotfiddle{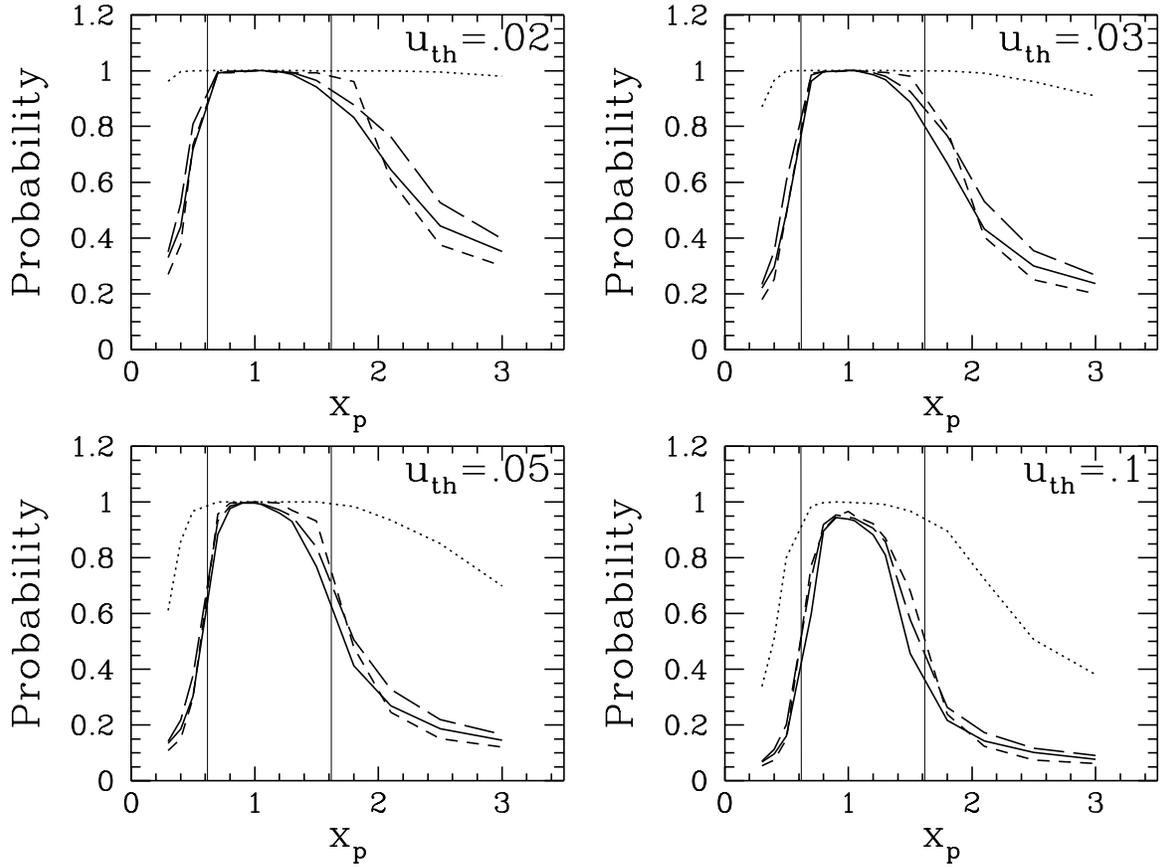}{7cm}{-90}{60}{60}{-230}{360}
\caption{
Same as Figure~\ref{figprobjup} except for $q=0.001$, corresponding
to a Saturn mass planet around a $0.3\msun$ star, or a Jupiter mass
planet around a $1\msun$ star.
\label{figprobsat}}
\end{figure}
\clearpage

\begin{figure}
\plotfiddle{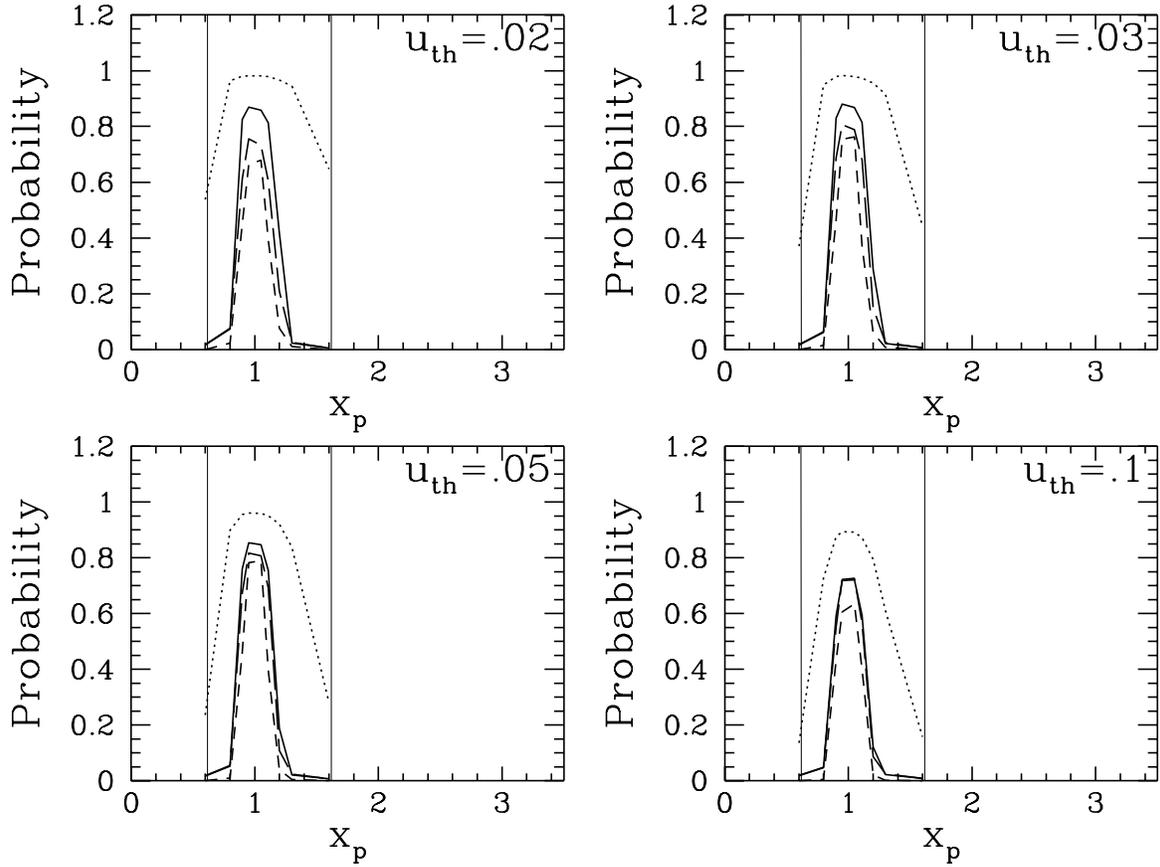}{7cm}{-90}{60}{60}{-230}{360}
\caption{
Same as Figure~\ref{figprobjup} except for $q=10^{-4}$, corresponding
to a 10 Earth mass or Uranus mass planet around a $0.3\msun$ star.
The source is $u_*=0.003$.
\label{figproburan003}}
\end{figure}
\clearpage

\begin{figure}
\plotfiddle{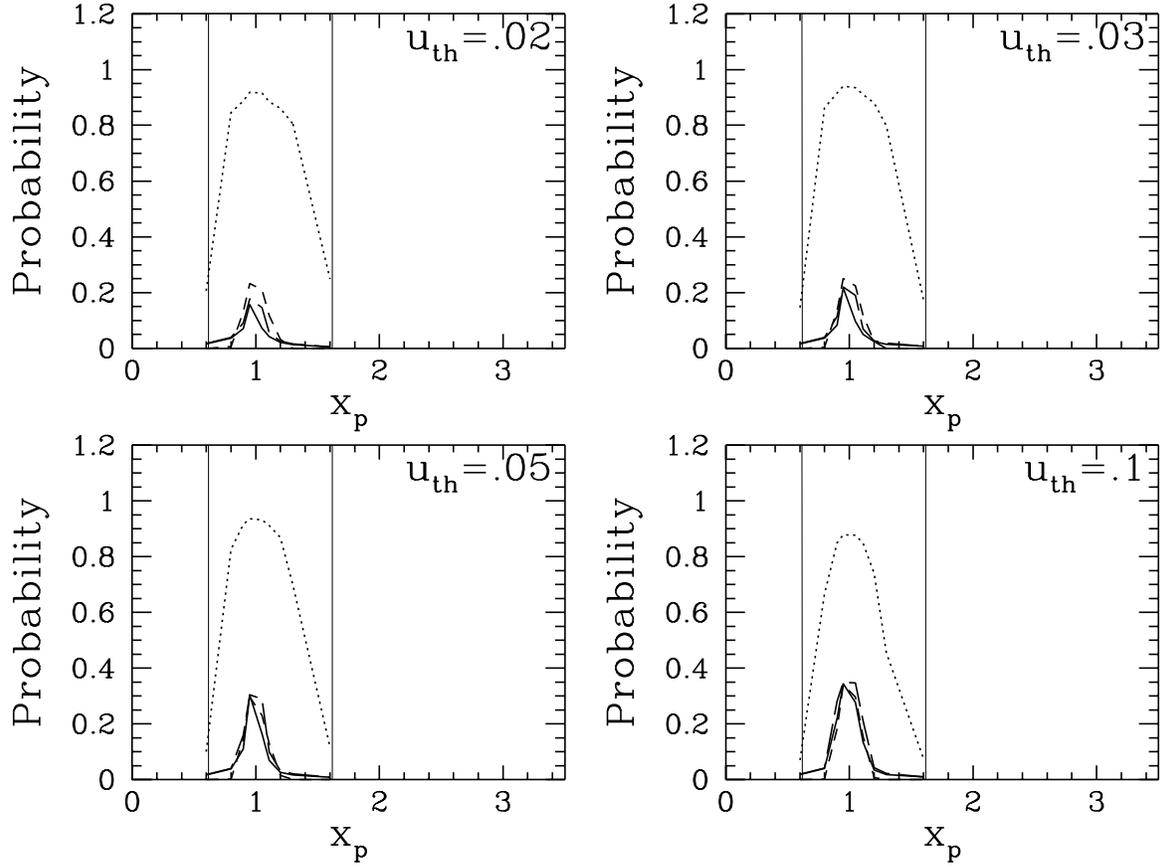}{7cm}{-90}{60}{60}{-230}{360}
\caption{
Same as Figure~\ref{figproburan003} except that the source radius is
$u_*=0.01$, corresponding to a giant source star. 
\label{figproburan01}}
\end{figure}
\clearpage

\begin{figure}
\plotfiddle{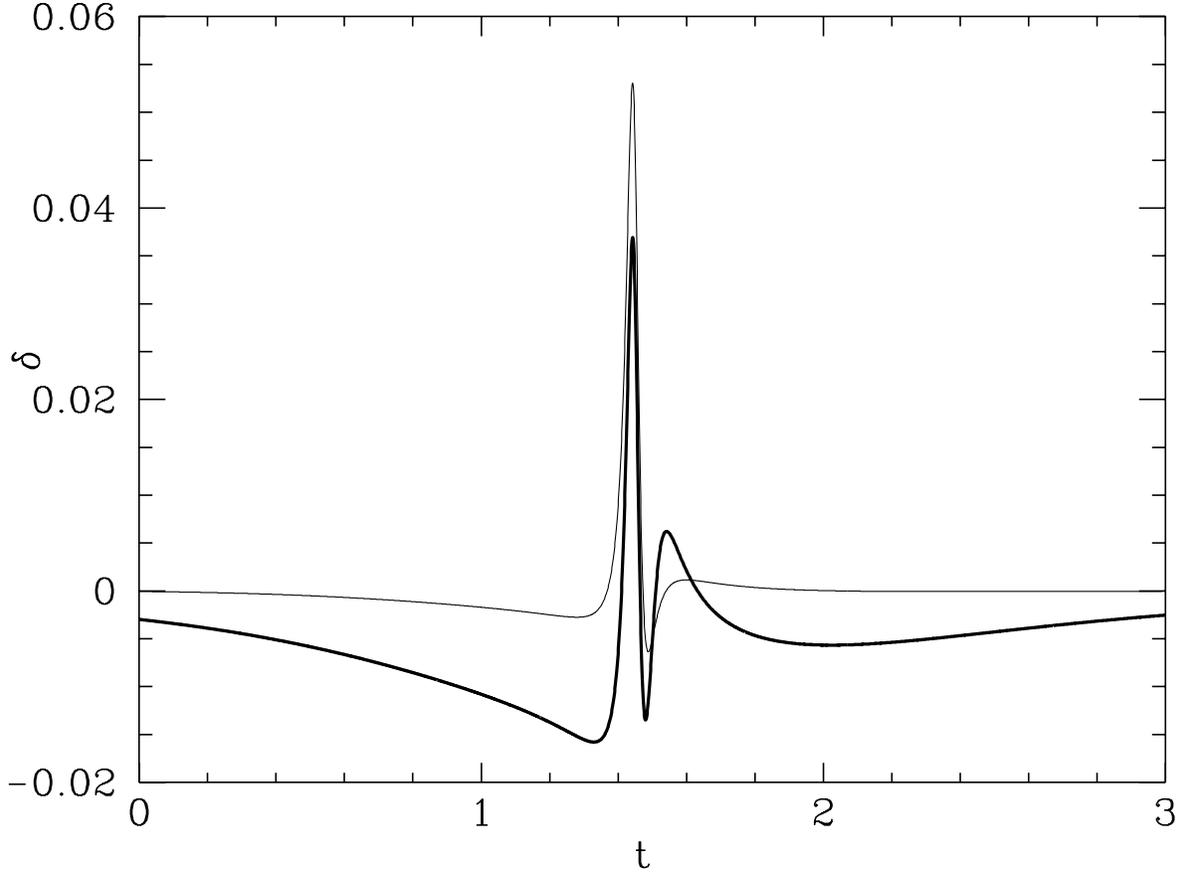}{7cm}{-90}{60}{60}{-230}{360}
\caption{
Comparison of planetary deviation lightcurve when using fitting to
find the single-lens subtraction lightcurve, 
and when using the known single-lens
lightcurve.  The quantity $\delta = (A_{bin}-A_{single})/A_{single}$ is plotted
vs. the time in units of $t_E$.  The light line uses the known single
lens parameters in the $\delta$ subtraction, while the heavy line finds
the best-fit single lens parameters from the binary lightcurve.
The fit lightcurve is less likely to be detected
when threshold statistics are used.  Parameters are $q=0.001$,
$x_p=1.5$, $\umin=0.05$, and $\beta=50^0$.
\label{figfitlc}}
\end{figure}

\end{document}